\documentclass[pre,aps,twocolumn,superscriptaddress,floatfix]{revtex4}

\usepackage{bm}
\usepackage{graphics}

\newcommand{\tauxb}{\tau_\mathrm{x}^{(0)}}
\newcommand{\taupb}{\tau_\alpha}

\newcommand{\fig}[4]
{
\begin{figure}[#4]
\resizebox{8.0cm}{!}{\includegraphics{#1}}
\caption{\label{#3}#2}          
\end{figure}
}

\newcommand{\figfig}[7]
{
\begin{figure}[#7]
\resizebox{#2}{!}{\includegraphics{#1}}
\resizebox{#4}{!}{\includegraphics{#3}}
\caption{\label{#6}#5}          
\end{figure}
}

\newcommand{\figfigstar}[7]
{
\begin{figure*}[#7]
\resizebox{#2}{!}{\includegraphics{#1}}
\resizebox{#4}{!}{\includegraphics{#3}}
\caption{\label{#6}#5}          
\end{figure*}
}

\begin{document}

\title{Negative differential mobility of weakly driven particles in models of glass formers}

\author{Robert L. Jack}
\affiliation{Department of Chemistry, University of California, Berkeley, CA 94720, USA}
\affiliation{Department of Physics, University of Bath, Bath BA2 7AY, UK}

\author{David Kelsey}
\affiliation{Department of Chemistry, University of California, Berkeley, CA 94720, USA}

\author{Juan P. Garrahan}
\affiliation{School of Physics and Astronomy, University of Nottingham,
             Nottingham NG7 2RD, UK}

\author{David Chandler}
\affiliation{Department of Chemistry, University of California, Berkeley, CA 94720, USA}

\begin{abstract}
We study the response of probe particles to weak constant driving in kinetically constrained models of glassy systems, and show that the probe's response can be non-monotonic and give rise to negative differential mobility:  increasing the applied force can reduce the probe's drift velocity in the force direction.  Other significant non-linear effects are also demonstrated, such as the enhancement with increasing force of the probe's fluctuations away from the average path, a phenomenon known in other contexts as giant diffusivity. We show that these results can be explained analytically by a continuous-time random walk approximation where there is decoupling between persistence and exchange times for local displacements of the probe.  This decoupling is due to dynamic heterogeneity in the glassy system, which also leads to bimodal distributions of probe particle displacements.  We discuss the relevance of our results to experiments.
\end{abstract}

\maketitle

\section{Introduction}

Dynamic heterogeneity manifests complex correlated atomic motions in structural glass forming systems~\cite{DH}.  
It is a ubiquitous feature that gives rise to a variety of behaviors 
peculiar to glassy dynamics.  In this paper, we consider 
phenomena associated with one such behavior -- the negative 
response of a particle's velocity to an applied force. 
Earlier experiments and computer simulations have studied how 
particles in glass forming systems respond to external 
forces~\cite{Weeks04,Reichhardt03,Evans06,forceInt,Sellitto}.  
A range of effects have been found, including the appearance of a 
threshold force~\cite{Weeks04} and 
non-linear velocity/force relations~\cite{Weeks04,Reichhardt03}, 
the vanishing of the linear response regime at low 
temperatures~\cite{Evans06}, the self-organisation of forced 
particles~\cite{forceInt}, and negative response to chemical potential gradients \cite{Sellitto}. Here, we study the response of a 
probe particle~\cite{YJ04} to a weak external force with numerical 
simulations and with analytic methods.  The numerical work employs 
kinetically constrained models of glass formers~\cite{FA84,Ritort-Sollich}.  
The analytical work employs a continuous time random walk 
model~\cite{Montroll}, specifically the model~\cite{tauk} introduced 
to treat effects due to decoupling between persistence and exchange 
processes~\cite{YJ04, YJ05}.  While originally constructed in the context of kinetically constrained lattice models, essential features of dynamics that justify this analytical model have been demonstrated in atomistic models of glass formers~\cite{chaudhuri,Hedges, Heuer}.  

\fig{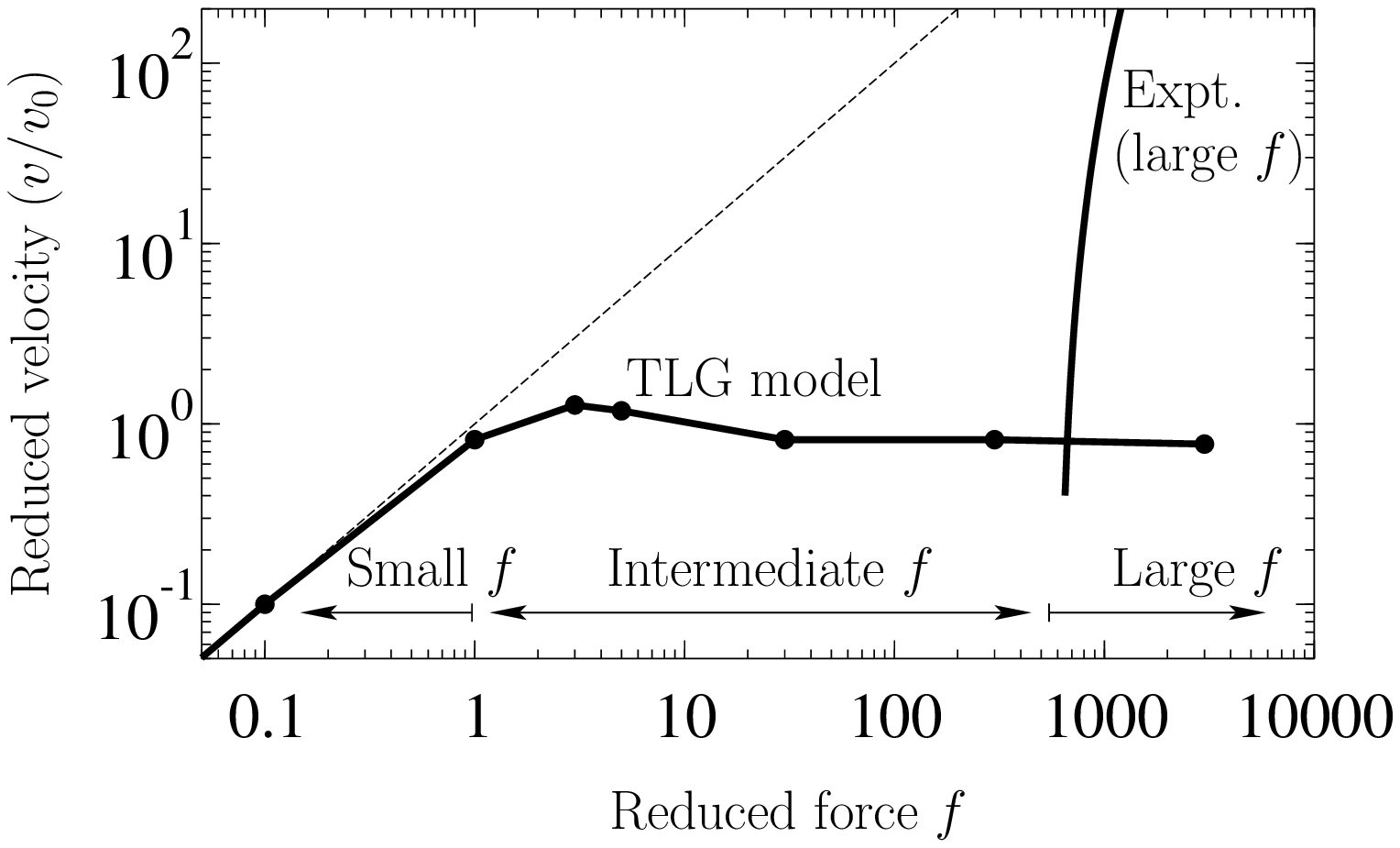}{
{\bf Experimental context}.  We plot
force-velocity relationships in rescaled units, 
identifying three regimes, as discussed in the main text.
The linear response (Einstein) relation, $v=v_0 f$, 
is represented by the dashed line.  
At small forces, the predictions of linear response apply.
At large forces, the experimental system exhibits
a threshold force that represents the limit of
applicability of the TLG model.
In the intermediate force regime, both the experimental and
TLG velocities 
are much smaller than those predicted by linear response,
and the response in the TLG model is non-monotonic.
The experimental curve shows the behaviour observed
in the colloidal fluid of Ref.~\cite{Weeks04}: we
plot the fitting function
 $v=v_\mathrm{t}[(F/F_\mathrm{t})-1]^\alpha$ used 
in~\cite{Weeks04}, with $F_\mathrm{t}=0.6\,
\mathrm{pN}$, $v_\mathrm{t}=2\times10^{-1}
\mu\mathrm{m}\,\mathrm{s}^{-1}$, and $\alpha=2.5$.  
For the experimental data, a reduced force of unity 
corresponds to $(k_\mathrm{B}T/\sigma_\mathrm{probe})\simeq 1\,
\mathrm{fN}$, and $v_0=(D/\sigma)\simeq10^{-5} \mu\mathrm{m}\, 
\mathrm{s}^{-1}$. The data for the TLG model was
 obtained at filling fraction $\rho=0.6$.
}
{fig:vf_exp}{ht}

Experimental context for our results is shown in Fig.~\ref{fig:vf_exp}:
we compare results from one of our model systems with results for
a colloidal system near to its glass transition.
In order to make contact between model systems and experiments, we plot 
the probe velocity $\bm{v}$ and the applied force $\bm{F}$ in dimensionless 
units.  The velocity is normalised by $v_0=(D/\sigma)$ where 
$\sigma$ is the particle
diameter, and $D$ is the diffusion constant of the probe in the absence of the externally applied force.  The reduced velocity $|\bm{v}|/v_0$ is proportional to the `modified Peclet number' of Ref.\cite{Weeks04}. 
The reduced force is $\bm{f}=(\bm{F}\sigma/k_\mathrm{B}T)$, 
where $T$ is the temperature and $k_\mathrm{B}$ is Boltzmann's constant.  
In these units, the linear response formula for the probe velocity 
is Einstein's relation for small forces, 
$\bm{v}=v_0\bm{f}$~\cite{Hansen-book}.
Estimating these reduced quantities using data from 
Refs.~\cite{Weeks04,Weeks02, vel-footnote}, we sketch the 
force-velocity relationship for the colloidal system in 
Fig.~\ref{fig:vf_exp}, where $v=|\bm{v}|$ and $f=|\bm{f}|$.  

The experimental data of Ref.~\cite{Weeks04} has several
important features.  The most striking is the so-called
threshold force, above which the velocity increases
rapidly with applied force.  On the other hand, for small enough
forces, the Einstein relation must apply, although
experimental constraints meant that this regime was
not accessible in Ref.~\cite{Weeks04}.  Interestingly,
for the experimentally accessible forces below the threshold, the velocity
is much smaller than the prediction of linear response:
in fact, it was smaller than the experimental resolution limit.
We can therefore identify a regime of intermediate force,
where the force is much smaller than the threshold, and the
response is much smaller than that predicted by the Einstein relation.

According to our theoretical predictions, this regime, which has
not yet been investigated experimentally, exhibits
new and interesting phenomenology.  
Figure~\ref{fig:vf_exp} illustrates our predictions by showing results from our simulations of a kinetically constrained model (details are given below).  
At small forces, the Einstein relation is obeyed, while the response 
saturates at larger forces.
This saturation represents a non-linear response that is consistent
with the small sub-threshold responses observed in Ref.~\cite{Weeks04}.
Our use of kinetically constrained models to describe the colloidal
fluid rests on the assumption that
glassy behaviour occurs when the motion
of particles is constrained by their neighbours.  We will see
that saturation of the sub-threshold responses is a natural
consequence of this assumption.  However, it is clear
from Fig.~\ref{fig:vf_exp} that the TLG model does not
reflect the experimental observation of a threshold force.
Our interpretation of the experimental data is that,
as the applied force is increased through the threshold, the force 
on the probe particle becomes stronger than
the constraint forces imposed by its neighbours. 
As a result, the structure of
the colloidal fluid is disrupted by the forced probe.  Thus
the threshold force represents the limit of applicability
of the kinetically constrained model~\cite{inviolable-foot}.  
For this reason, we concentrate in this article on the regime of 
intermediate forces, where the kinetically
constrained model should be applicable. In this regime, we find
a surprising effect: increasing the applied force \emph{reduces} the 
velocity of the pulled particle.  This phenomenon and associated results
are the focus of this paper.  

\section{Models and particle driving protocols}
We begin by describing the models we use in numerical simulations.  Two are ``particle'' models, in the sense that the material in which the probe moves is described in terms of particles on a lattice with specified dynamical rules.  The other is a ``field'' model in the sense that the material is described in terms of a so-called ``mobility'' field.  The value of that field at a point on the lattice specifies whether or not motion is possible at that point.  The specific field model is one of many possibilities, each representing a coarse grained approximation to a particle model~\cite{JPH&DC'03}.  
The continuous-time random walk model~\cite{tauk} used later in this paper 
describes the coupling of the mobility field to particle motion.

\subsection{Particle models}

We consider a two-dimensional (i.e., square) version of Kob and Andersen's 
(KA) lattice model~\cite{KA93,Toninelli-KA}, 
and the closely related triangular lattice gas (TLG) model of J\"{a}ckle and 
Kr\"{o}nig~\cite{JackleTLG,Pan-tlg}.
These constrained lattice gases are assumed to capture effects of local
jamming in a supercooled liquid.  They have only excluded volume interactions,
so, for a given filling
fraction $\rho$, all allowed configurations are equally likely.  
They exhibit non-trivial effects of correlated dynamics 
at filling fractions $\rho \gtrsim 0.5$.  
Dynamical heterogeneity is manifested as a clustering
of mobile particles, and different transport properties decouple
from one another (for example, the translation
diffusion coefficient does not scale inversely with the
structural relaxation time~\cite{Pan-tlg}).
These effects arise from constraints on the dynamics.  

For the KA model \cite{KA93,Toninelli-KA}, specifically the (2,2) variant, particles live on a square lattice, with zero or single occupancy of lattice sites, and a particle may move to adjacent vacant sites
only if the particle is adjacent to two vacant sites in both the 
initial and final sites.  Similarly, for J\"{a}ckle and Kr\"{o}nig's
TLG model, there is single or zero occupancy on a triangular lattice,  and a particle may move to vacant sites only if both of the
mutual nearest neighbours of the initial and final sites are 
vacant.  Accordingly, we refer to this model as the (2)-TLG~\cite{Pan-tlg}.
Its dynamical rules can be motivated on physical
grounds by associating hard-cores to the particles, the diameter of which is equal to the lattice spacing, and by insisting that particles move along the lines connecting the points on the lattice.  From that picture, it is seen that the dynamical rule is a straightforward steric effect -- there is not enough room for a particle to pass to an empty nearest neighbor site unless the two common adjacent sites are also vacant. 

In the absence of applied forces, all allowed processes happen with rate $\gamma$, which sets the fundamental unit of time.
We implement the dynamics using asynchronous Monte Carlo updates.

In order to introduce a force $\bm{F}$ on a single ``probe'' particle, we
first consider that particle in an empty lattice.  It would
correspond to a single colloidal particle alone in a solvent.
In the absence of the applied force, this particle has a bare diffusion
constant $(\sigma^2\gamma z/2d)$ where $z$
is the co-ordination number of the lattice, and we identify the
particle diameter $\sigma$ with the lattice spacing.
We denote the force-dependent rates for translational moves of 
displacement $\Delta\bm{R}$ by $W(\Delta\bm{R},\bm{F})$.  
We assume that these rates
obey a local form of detailed balance, 
$W(\Delta\bm{R},\bm{F})={\rm e}^{\bm{f}\cdot\Delta\bm{r}} 
W(-\Delta\bm{R},\bm{F})$,
where $\Delta\bm{r}=\Delta\bm{R}/\sigma$ is a reduced displacement.
To ensure that an isolated probe obeys the Einstein 
relation for all forces, we use the rates
\begin{equation}
W(\Delta\bm{R},\bm{F})=2\gamma
        \frac{ g(\bm{f}\cdot\Delta{\bm{r}})  }
             {1+\exp(-\bm{f}\cdot\Delta{\bm{r}})},
\label{equ:rates}
\end{equation}
where the dimensionless function $g(E)$ is given by
\begin{equation}
g(E) \equiv E/[2\tanh(E/2)] .
\label{g} 
\end{equation}
We have $g(0)=1$, so that $W(\Delta\bm{R},\bm{0})=\gamma$.
An alternative choice would be to use
Glauber dynamics [$g(E)=1$ for all $E$], but in that case,
the Einstein relation applies only for
small velocities, and  there is an unphysical saturation 
at large forces.
This saturation is an artifact of using a lattice to describe 
continuous space~\cite{latt-foot}: by using the rates
of Equ.~(\ref{equ:rates}), we ensure that the velocity
of a single particle does not saturate. 
Hence, the saturating response that we do observe is a physical
effect, which appears as the motion of the
probe particle becomes increasingly constrained by other
particles in the system.
We have also checked that while the data presented in this 
article do depend quantitatively on the choice of $g(E)$, 
their qualitative features are preserved if we instead
use Glauber dynamics.  
Thus, while there is some arbitrariness
in our use of Eq.~(\ref{equ:rates}), we are confident
that this choice does not affect our main conclusions.

Throughout this article, we consider a single probe 
in a large system of unforced particles, 
ignoring the collective behaviour arising from interactions
between forced particles~\cite{forceInt,Sellitto}, and the effect
of finite current densities~\cite{Sellitto}.  To enhance 
statistics, we simulate large systems with a few probe particles in each, 
and we verify that our results are independent of the number of probes.
Due to the underlying lattice, our models are not isotropic, 
so there is some dependence on
the direction of the force compared to the lattice axes 
(for example, see \cite{Reichhardt03}). 
However, we find that
the results are qualitatively similar for all angles.

\subsection{Mobility field model}

We also consider the one-dimensional one spin facilitated Fredrickson-Andersen
(FA) model \cite{FA84,GC02, BG03}.  In this model the 
local structure is described by a binary
variable $n_i\in\{0,1\}$.  Sites with $n_i=1$ are `mobile', 
or excited:
particles in these regions are able to move; those with $n_i=0$ 
are `jammed', so that motion is very unlikely. 
Sites may change their state only if they are adjacent to a site 
with $n_i=1$. 
In that case they flip from $0$ to $1$ with rate $c=e^{-\beta}$
and from $1$ to $0$ with rate unity (this choice sets the unit
of time in this model), where $\beta$ is a dimensionless
inverse temperature.  The dynamics
obey detailed balance, with an energy function $E=\sum_i n_i$,
so the equilibrium state has no correlations between sites.  This model exhibits effects of non-trivial correlated dynamics for $\beta \gtrsim 1$.

While the relaxation time of the FA model exhibits Arrhenius 
temperature dependence~\cite{Ritort-Sollich}, in $d=1$ and at
low temperature, the model has significant fluctuation effects.
These are manifested by stretched exponential 
relaxation~\cite{Ritort-Sollich} and transport decoupling~\cite{YJ04}.
Other kinetically constrained models of mobility fields, such as the
East model~\cite{Ritort-Sollich,Jackle-East} exhibit similar phenomenology 
in all dimensions, but the $1d$ FA model is sufficient to capture
the essential physical effects discussed in this article~\cite{East-foot}.

In Ref.~\cite{YJ04}, probe particles were coupled to the FA model.  Probes were
allowed to hop between adjacent sites only when both initial and
final sites were mobile.  With this choice, the dynamical
rules for the $n_i$ do not depend on the positions of the probe particle.
While the forced probe is affected by its environment, there is no 
mechanism for a back-reaction, where the effect of the force
feeds back onto the $n_i$ variables.
Thus, an excitation in an FA model can pass through this type of probe particle
without any knowledge of the probe's presence.  We therefore refer to the 
probes of Ref.~\cite{YJ04} as `ghost' probes. 

For particle-based models, including the KCMs introduced above,  
forced probe particles do have significant
effects on their environment.  For
this reason we introduce an alternative set of rules for probe
particles in the FA model, which allow us to model this back-reaction.  In particular, since the presence of a probe reduces the available free volume, we imagine that its presence is sufficient to render the site immobile.  Thus, we modify the model of Ref.~\cite{YJ04} by assuming probes cannot occupy mobile sites.  In this case, the probe moves through the system by swapping places with mobility excitations.  If the probe position is $x$, then the allowed moves are
$$
(x=i,n_i=0,n_{i\pm1}=1) \to (x=i\pm1 ,n_i=1,n_{i\pm1}=0)
$$
Clearly, the dynamics of the FA
model itself are now coupled to those of the
probes.  We therefore refer to these as `fully coupled probes'.
The rates for attempted moves of both coupled and ghost probes
are given by Eq.(\ref{equ:rates}), with $\gamma=1/2$. 

\section{Force-dependence of the probe velocity}

\fig{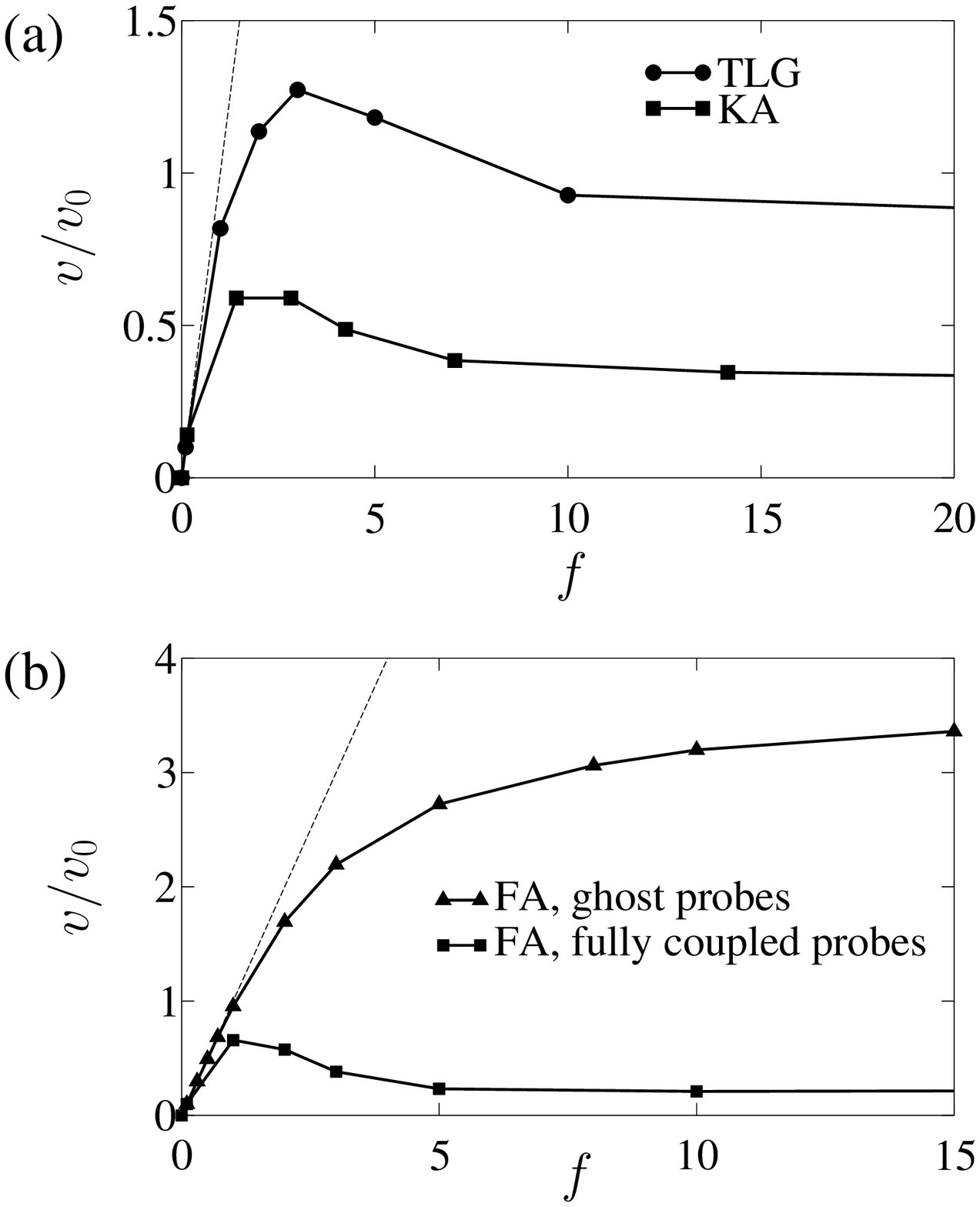}{
(a) Non-monotonic velocity-force relationships in particle
models, illustrating negative differential mobility.  The
data for the J\"{a}ckle-Kr\"{o}nig TLG model were obtained at
filling fraction $\rho=0.6$, with a force applied along a lattice axis.
In the corresponding KA model, the force was applied at $45^\circ$ to the axes, 
at filling fraction $\rho=0.88$.  The long time limit
in the definition of $\bm{v}$ necessitates the use of 
long trajectories (typically $10^4\tau_\alpha$).
(b) Comparison of ghost and fully coupled probes
in the FA model in one dimension, at $\beta=3$.  Only the fully
coupled probes show a negative differential mobility, indicating
that this effect arises from a reaction of the mobility field to
the forced probe. 
%
}{fig:vf_model}{ht}

\fig{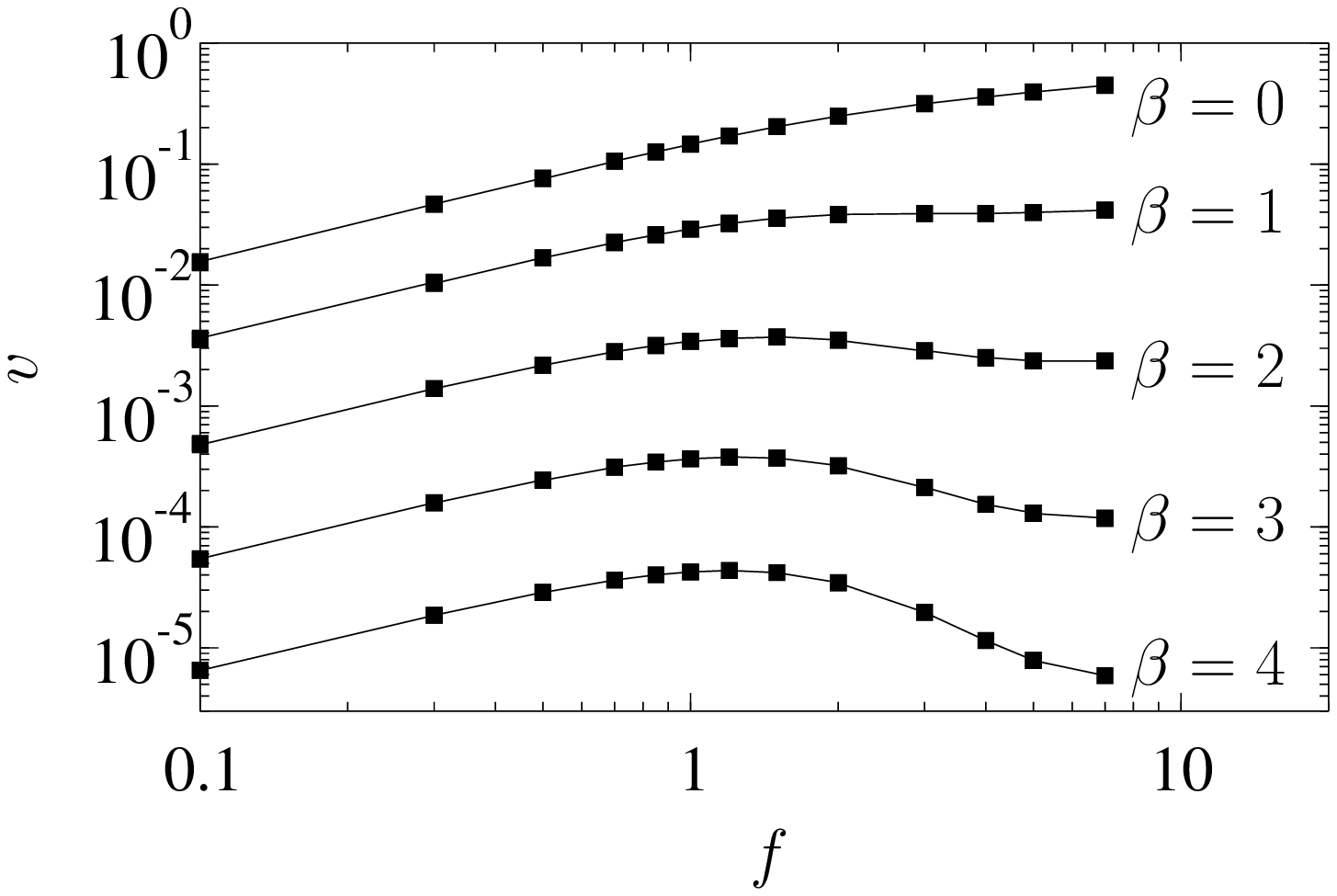}{
Data showing the temperature dependence of the velocity-force
relationship in the FA model with fully-coupled probes.  
Non-monotonic response occurs in the low temperature
regime $\beta\gtrsim1$.
}{fig:vf_temp}{ht}

\subsection{Negative differential mobility}
\label{sec:neg}

\figfigstar{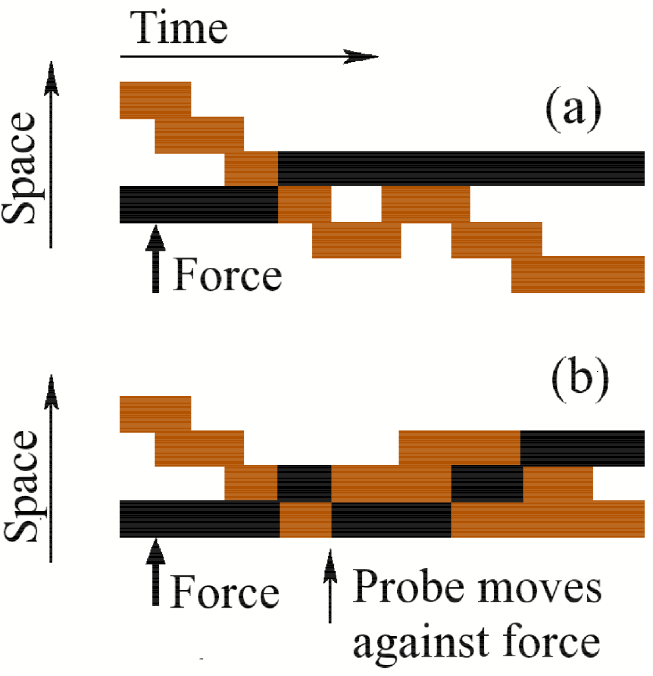}{4cm}{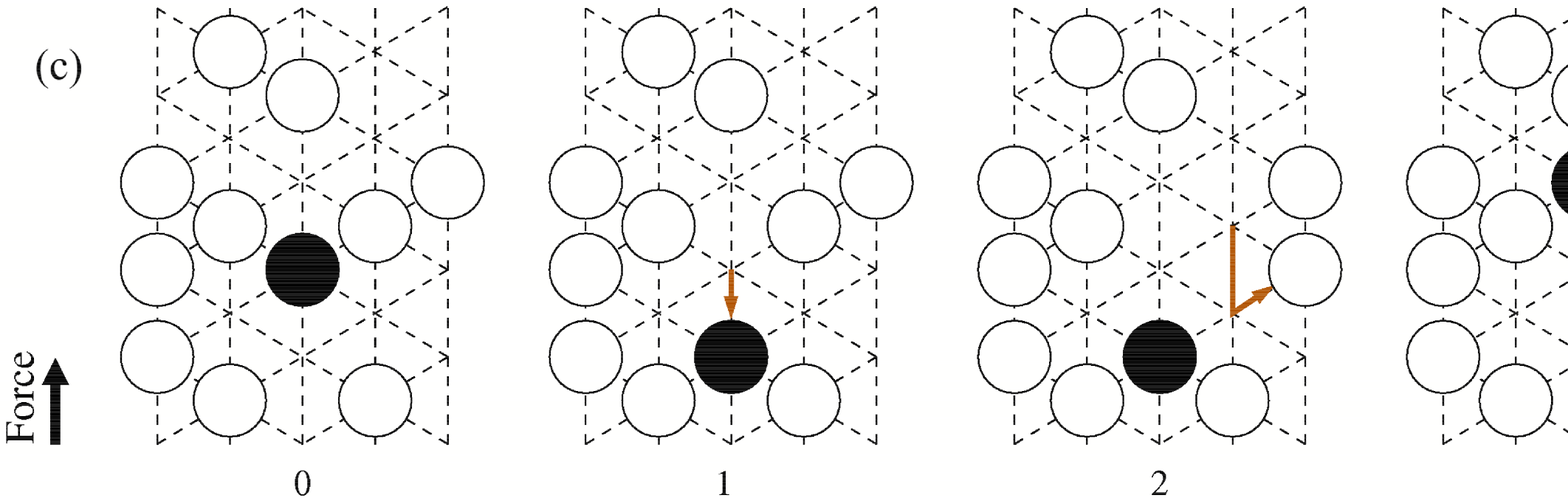}{13cm}{
(Color online) Sample mechanisms for reduced velocity at large forces.
(a,b) Sketch of two trajectories in the FA model. 
Excited sites, $n_i=1$, are shown in gray (or brown), the trajectory of 
the probe in black, and time evolves from 
left to right.  (a) A probe moves a single step
on encountering an excitation line.   
(b) An excitation line can facilitate several hops
for the probe, increasing its diffusion constant. 
However, this mechanism requires steps in which the probe
moves against the force, so it is suppressed for large
forces.  (c) A sequence of four configurations
that illustrate a co-operative move in the (2)-TLG.
The probe (coloured black) responds to the force by eventually moving 
upwards, but this requires an initial step
in which it moves downwards, against the force, to allow the neighbouring particles which are blocking it to move out of the way. Thus the response
is again suppressed at large $F$. 
}{fig:mech}{ht}

The drift velocity of the probe is
\begin{equation}
\bm{v}=\lim_{t\to\infty}t^{-1} \langle \Delta\bm{r}_\mathrm{probe}(t)\rangle,
\label{drift}
\end{equation}  
where $\Delta\bm{r}_\mathrm{probe}(t)$ is the displacement
of the probe in time $t$, and the average is taken in the
presence of the applied force. In Fig.\ \ref{fig:vf_model}(a) we plot this
velocity as a function of force in the 
(2)-TLG and (2,2)-KA models, at densities above their onsets to cooperative dynamics.  The linear response (i.e., Einstein) relation is $\bm{v}=(D/\sigma)\bm{f}$,
where $D$ is the zero-force diffusion constant,
\begin{equation}
D=\lim_{t\to\infty} (2dt)^{-1} 
\langle |\Delta\bm{r}_\mathrm{probe}(t)|^2\rangle_{f=0}.
\label{D }
\end{equation}
At densities above the onset, $\rho \gtrsim 0.5$, $D$ is much smaller than its bare value $(\sigma^2\gamma z/2d)$.  The striking result of negative response, i.e., negative differential mobility, $\mathrm{d}v/\mathrm{d}f<0$, is found for $f \gtrsim 2$. 

A similar effect is shown in Fig.\ \ref{fig:vf_model}(b) for the fully
coupled probes in the FA model. However, negative differential mobility does not occur for the ghost probes.  In that case, non-linear effects only lead to 
a saturation of the drift velocity with force~\cite{Ghost}.  
In Fig.~\ref{fig:vf_temp}, we show how this response depends on the 
temperature.  Keeping the reduced force $f$ fixed, the velocity
decreases monotonically as the temperature is reduced.  At high
temperatures, the dependence on the force is monotonic; 
negative differential mobility
appears below the onset temperature, $\beta\gtrsim 1$.  For
these temperatures, the linear response (small force) regime is 
$f\lesssim1$, and negative differential mobility is observed 
near $f=1$, below the onset temperature.

The lack of non-monotonic behavior for ghost particles shown
in Fig.~\ref{fig:vf_model}(b)
points to the mechanism for non-monotonic behavior.  We use Fig.\ \ref{fig:mech} to illustrate the mechanism. In particular, we show how applying a large force to a probe particle can prevent the movement of neighbouring
particles or mobility excitations.  For example,
Fig.\ \ref{fig:mech}(c) shows a configuration in the (2)-TLG for which
the probe can make progress in the direction of the force, but
only if it first moves backwards,
allowing the neighbouring particles to move out of its way.  
In the FA model, Fig.\ \ref{fig:mech}(b), illustrates how
a single excitation may allow a single probe to make
several steps along the direction of the force~\cite{YJ04}.  However, this
process is suppressed at large forces, since it also involves steps
in which the probe moves against the force.  At large forces, most
encounters between probe and excitation are of the form shown
in Fig.~\ref{fig:mech}(a), and the probe makes only one step
in each such encounter.  In both cases, 
the forced probe acts to suppress local relaxation, and the force
acts to slow down the motion of the probe particle.

The results of Ref.\ \cite{Sellitto} for KA models show a non-monotonic response similar to that of Fig.~\ref{fig:vf_model}.  In that case, all particles attain a finite drift velocity due to force gradients applied to many particles, and large scale density changes result.  The effects found in that work are related to those we present here, but in contrast to Ref.\ \cite{Sellitto}, the effect we consider arises from forces on a single forced particle that produce no large scale density changes.  Non-monotonic response of a drift velocity has also been demonstrated for systems with quenched disorder~\cite{Cecchi96-others}, but the origin of these phenomena is different from those that underlie the results of Ref.\ \cite{Sellitto} and of Fig.~\ref{fig:vf_model}, where the mechanisms involve the dynamics of the medium.

Finally, we note that the saturation velocity is finite in all
of the models that we consider.  This is to be contrasted with
systems with large numbers of forced particles, in which 
their velocity may appear to vanish at large forces~\cite{Sellitto}.
For single forced particles, our results suggest that the saturation
velocity will be finite as long as the unbiased diffusion constant 
is finite.  [In KCMs with glass transitions at finite 
densities (or temperatures)~\cite{ne-spiral}, all particles
are localised in the glass phase: we expect $D=0$ and $v(f)=0$ for
all $f$.]

\section{Continuous time random walk analysis}
\label{sec:ctrw}

To further elucidate the mechanism for the non-monotonic responses
described above, we use a continuous-time random walk (CTRW) 
analysis~\cite{Montroll}.  We use the model of~\cite{tauk}, 
which exploits the existence of two separate timescales 
in glassy systems, corresponding to two different physical processes:
exchange and persistence events~\cite{tauk,YJ05,Hedges}.
Recently, Rubner and Heuer~\cite{Heuer} analysed how motion on an underlying energy landscape can result in particle motion that resembles a CTRW. Earlier applications of CTRWs to glassy materials include trapping models such 
as~\cite{trap}, and mostly focussed on the physical consequences of diverging time scales.  In the following, we will assume that all time scales are finite, ensuring that we recover the physical limit of simple diffusion at long times.

\subsection{Theoretical framework}

Consider a single probe particle, which makes a series of uncorrelated steps through
a fluctuating environment.  The force on the probe enters as a bias on the direction of these steps.  The probe's environment is dynamically heterogeneous, with space-time permeated with entangled excitation lines~\cite{GC02, YJ04, YJ05}.  The probe moves only where it is intersected by these lines.  In this way, the time between successive steps acquires large fluctuations.  We denote the distribution of these times by $\psi(t)$, so that the distribution of time intervals between a randomly selected initial time and the first step made by a particle is \cite{renewal}
\begin{equation}
p(t) = \frac{\int_t^\infty dt' \psi(t') }{\int_0^\infty dt' t' \psi(t')}.
\label{equ:p-psi}
\end{equation}
Following~\cite{YJ04,YJ05}, we refer to $\psi(t)$ as the distribution of exchange times and $p(t)$ as the distribution of persistence times.  The probability that the particle has made no steps between time $0$ and time $t$, the so-called ``persistence'' function, is 
\begin{equation}
P(t)=\int_t^\infty\!\mathrm{d}t' p(t').
\label{equ:P}
\end{equation}

Now, let $G(\bm{r},t)$ be the distribution
of the position $\bm{r}$ of the probe particle, given that
it was at the origin a time $t$ earlier.  The number of hops
made by the probe in that time is randomly distributed, so 
$G(\bm{r},t)$ contains a contribution from every possible
number of hops.  Since successive hops
are assumed to be
independent, the Fourier-Laplace representation of the
sum over these contributions has a closed
form:
we define $\hat{F}(\bm{k},s)=
\int_0^\infty\mathrm{d}t  
\int\mathrm{d}^d\bm{r}\, G(\bm{r},t) e^{-i\bm{k}\cdot\bm{r}-st}$,
and arrive at the Montroll-Weiss equation~\cite{Montroll,tauk}
\begin{equation}
\hat{F}(\bm{k},s) = \hat{P}(s) + \frac{1-\hat{\psi}(s)}{s} 
     \frac{\Gamma(\bm{k})}{1-\hat{\psi}(s) \Gamma(\bm{k})} \hat{p}(s).
\label{MW}
\end{equation}
The functions $\hat{\psi}(s)$, $\hat{P}(s)$ and $\hat{p}(s)$ 
are the Laplace transforms of 
$\psi(t)$, $P(t)$ and $p(t)$, respectively.  The function $\Gamma(\bm{k})$ is the generating function for the statistics of a single random walk step, $\Delta \bm{R}$,
\begin{equation}\Gamma(\bm{k}) \equiv \overline{e^{-i\bm{k}\cdot
\Delta \bm{R}}},\end{equation}
where the overbar indicates an average over the possible values of $\Delta \bm{R}$.  For small wave-vectors,
$\Gamma(\bm{k}) = 1 - i\sigma\bm{\delta}\cdot\bm{k} - (\sigma^2/2)|\bm{k}|^2 + \mathcal{O}(k^3)$, where the invariance of the quadratic term under rotation ensures that unforced diffusion is isotropic, while in the presence of a force there is a non-trivial bias $\bm{\delta} = \overline{\Delta\bm{R}}/\sigma$.  The length scale $\sigma$ would refer to a particle diameter in continuous force models, and here refers to the lattice spacing of our KCMs.

The behaviour of the probe particle at long times is given by the 
behaviour of $\hat{F}(\bm{k},s)$ at small $k$ and $s$.  
We denote the mean exchange and
persistence times by $\tau_\mathrm{x}$
and $\tau_\alpha$ respectively 
(we identify the average persistence time of the probe with the structural 
relaxation time $\tau_\alpha$
of the embedding medium~\cite{alpha-foot}).
From (\ref{equ:p-psi}), we have $\hat{\psi}(s)=1-s\tau_\mathrm{x} \hat{p}(s)$,
and we expand at small $s$, arriving at
$\hat{p}(s) = 1 - s \tau_{\alpha} + \dots$
and $\hat{\psi}(s) = 1 - s \tau_{\rm x} + s^2 \tau_\mathrm{x}
\tau_\alpha + \dots$.  

The drift velocity is then
\begin{equation}
\bm{v} = \lim_{s\to0} is^2 \nabla_{\bm{k}} \hat{F}(\bm{k},s)|_{k=0}
 = (\sigma / \tau_\mathrm{x}) \bm{\delta},
\label{rs}
\end{equation}
where we used $\bm{v}=\lim_{t\to\infty} 
t^{-1} \int \mathrm{d}\bm{r}\, \bm{r} G(\bm{r},t)$.
Physically, the drift velocity is given by the product of the mean
displacement per hop, $\sigma\bm{\delta}$, and the mean hop
frequency, $\tau_\mathrm{x}^{-1}$.

Similarly, the asymptotic mean square fluctuation in the probe displacement per unit time
is given by
\begin{eqnarray}
D(\bm{f}) 
 &=& (2d)^{-1}\lim_{t\to\infty} t^{-1} 
   \left[ \langle |\bm{r}(t)|^2 \rangle - |\langle \bm{r}(t) \rangle|^2 \right]
  \nonumber \\
  &=& (2d)^{-1}\lim_{s\to0} \left[ -s^2 
      \nabla_{\bm{k}}^2 \log F(\bm{k},s)|_{k=0} - (|\bm{v}|^2/s) \right]
\nonumber \\
  &=& (2d\tau_\mathrm{x})^{-1} \left\{ \sigma^2+ 
  2 |\bm{\delta}|^2\sigma^2
   \left[ (\tau_{\alpha}/\tau_\mathrm{x})-1\right] 
\right\}.  \nonumber \\
\label{equ:mu2}
\end{eqnarray}
Hence, the unforced diffusion constant is
\begin{equation}
D=D(0)=\frac{\sigma^2}{2d\tau_\mathrm{x}}.
\label{equ:diff}
\end{equation}

\figfig{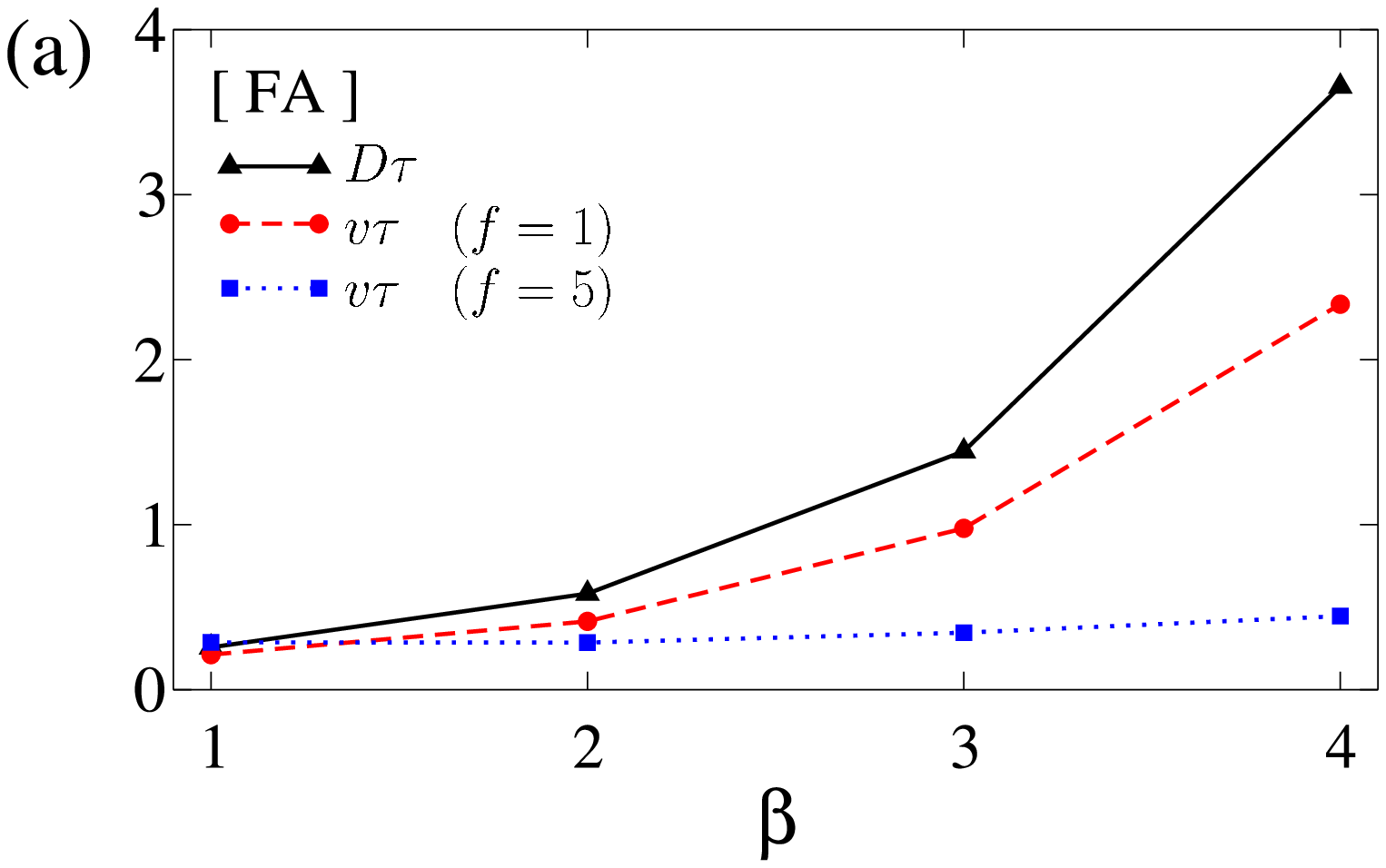}{8cm}{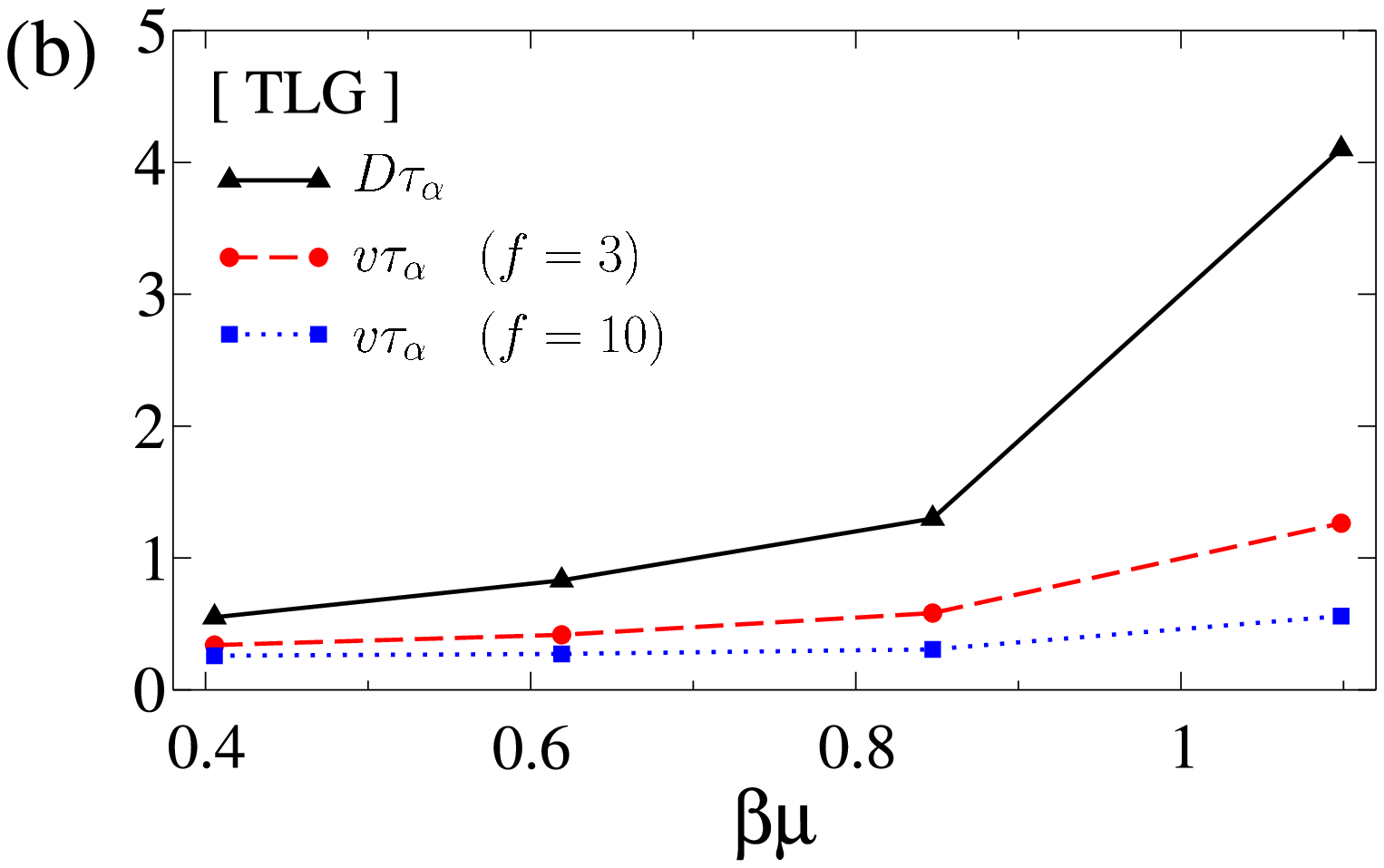}{8cm}{
(Color online) Scaling of the response with temperature and density.  
We plot rescaled transport coefficients 
as a function of inverse temperature or chemical potential. 
(a) FA model: we show the diffusion constant, 
the velocity at $f=1$ (close to the maximal response $v^*$)
and the velocity at $f=5$ (close to the saturation velocity 
$v_\mathrm{sat}$).  To investigate
the relative scalings of these quantities, we normalise them all
by the persistence time $\tau$ (which varies by around four orders
of magnitude across this temperature range).  
(b) (2)-TLG, for a range of filling fractions
$0.6\leq\rho\leq0.75$.  We normalise by the structural
relaxation time, which increases by a factor of around 300
across this range of density.  We define the chemical potential
for vacancies to be $\mu$, so that
$\rho=(1+e^{-\beta\mu})^{-1}$ increases from left to right.
The maximal
and saturation velocities are estimated using $f=3$ and $f=10$
respectively.  
}{fig:v_scale}{ht}

\subsection{Effect of forcing}
\label{sec:ctrw-neg}

From Eq.(\ref{rs}), we see that the probe drift velocity depends
on the bias $\bm{\delta}$ and the mean exchange time $\tau_\mathrm{x}$.  These quantities depend upon the force.  In one dimension, the local
form of detailed balance relates the probabilities
for hopping to left and right, and we have
\begin{equation}
\delta = \tanh(f/2).  
\label{equ:force-bias}
\end{equation}
This result is independent of the function $g(E)$ in Eq.(\ref{equ:rates}).
In $d>1$, the bias $\bm{\delta}$ is aligned with
the force, as long as diffusion in the unforced case is isotropic.
The modulus of $\bm{\delta}$ increases monotonically with the force, and
has a large-$f$ limit $0<\delta_\mathrm{max}\leq1$.

The dependence of the mean exchange time $\tau_\mathrm{x}$ on the force manifests the response of the medium on the probe. In the case of a ghost particle, there is no effect so that Eq.(\ref{rs}) gives (in one dimension)
\begin{equation}
[v(f)]_\mathrm{ghost} = \frac{\sigma}{\tauxb} \tanh(f/2),
\label{rtghost}
\end{equation}
where $\tauxb$ denotes the mean exchange time for the medium in the absence of the forced probe particle.  This drift velocity follows the Einstein relation for small forces, while for larger forces, it saturates at a limiting value of the order of $v_0$.  This is illustrated for the $d=1$ FA model in Fig.~\ref{fig:vf_model}(b).

More generally, the forced particle does affect its 
surroundings, and as discussed in 
Sec.~\ref{sec:neg} and Fig.~\ref{fig:mech}, applying a large
force to a probe particle tends to suppress the relaxation of the
surrounding medium.  In the language of the CTRW, this
local slowing down enters as an increase of the mean
exchange time.

In fact,
Fig.~\ref{fig:mech}(a) indicates that, for large forces in
the FA model, the probe
typically makes only one step for each excitation that it encounters.
For this model, the results of~\cite{YJ04,YJ05} indicate that
$\tauxb$ is the typical time between encounters
with the same excitation line, while $\tau_\alpha$ is the typical
time between encounters with different excitation lines.
Thus, if the mechanism of Fig~\ref{fig:mech}(a) is dominating,
we expect $\tau_\mathrm{x}\approx\tau_\alpha$
However, for small forces, the back reaction of the probe on the medium 
can be neglected and we expect $\tau_\mathrm{x}\approx\tauxb$.  
To interpolate between these two limits, we combine the
rates for the two processes using the simple functional form: 
\begin{equation}
\frac{1}{\tau_{\rm x}} \approx \frac{1 - |\delta|}{ \tauxb }
 + \frac{|\delta|}{\taupb} .
\label{tauxd}
\end{equation}
(We choose this form for simplicity, noting that analyses based
on it are mostly qualitative, and do not depend on the precise
form used.)
Using (\ref{tauxd}) in (\ref{rs}) we then obtain (for $d=1$),
\begin{eqnarray}
v(f) &\approx& (2D/\sigma)
  \tanh(f/2)
\left[
1 - \frac{\taupb-\tauxb}{\taupb} \tanh|f/2| 
\right] . 
\nonumber \\
\label{rtcoupled}
\end{eqnarray}
The velocity $v$ is non-monotonic in $f$, with a maximum at 
$f=\mathcal{O}(1) $.  
This is the non-monotonic behaviour of the 
drift velocity observed in Fig.\ \ref{fig:vf_model}.  
The peak value $v^*$ scales with $D$
while the large-force limit of the velocity, $v_\mathrm{sat}$,
scales with the inverse of $\tau_{\alpha}$.  Thus
we expect $v^*$ to scale with $D$, while $v_\mathrm{sat}\tau_\alpha$
depends only weakly on temperature and density.
Figure \ref{fig:v_scale}(a) shows that the behavior
of the fully coupled probe in the FA model is consistent with
this analysis.  (We have rescaled by the persistence time $\tau$
of the excitations $n_i$, which has the same scaling as the
mean persistence time of the probes $\tau_\alpha$~\cite{YJ04}.)
Generalizing (\ref{rtcoupled}) to $d>1$ leads to a similar prediction 
of non-monotonic response; the scaling of the maximal
and saturation velocities is shown in Fig.~\ref{fig:v_scale}(b), and
the qualitative features are again consistent with the CTRW analysis.

\subsection{Force-dependent fluctuations: Giant diffusivity}
\label{sec:fluct}

\figfig{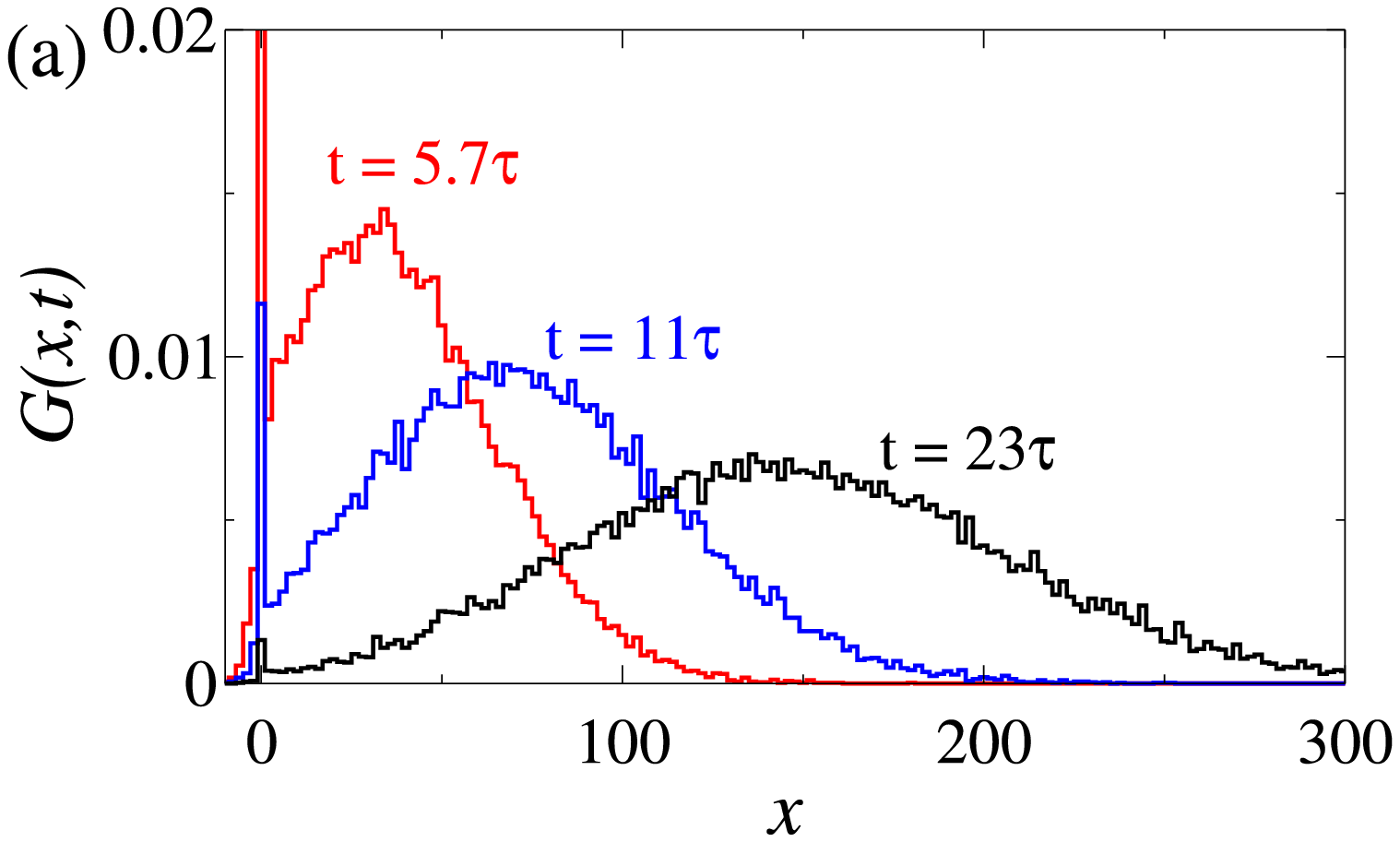}{8.5cm}{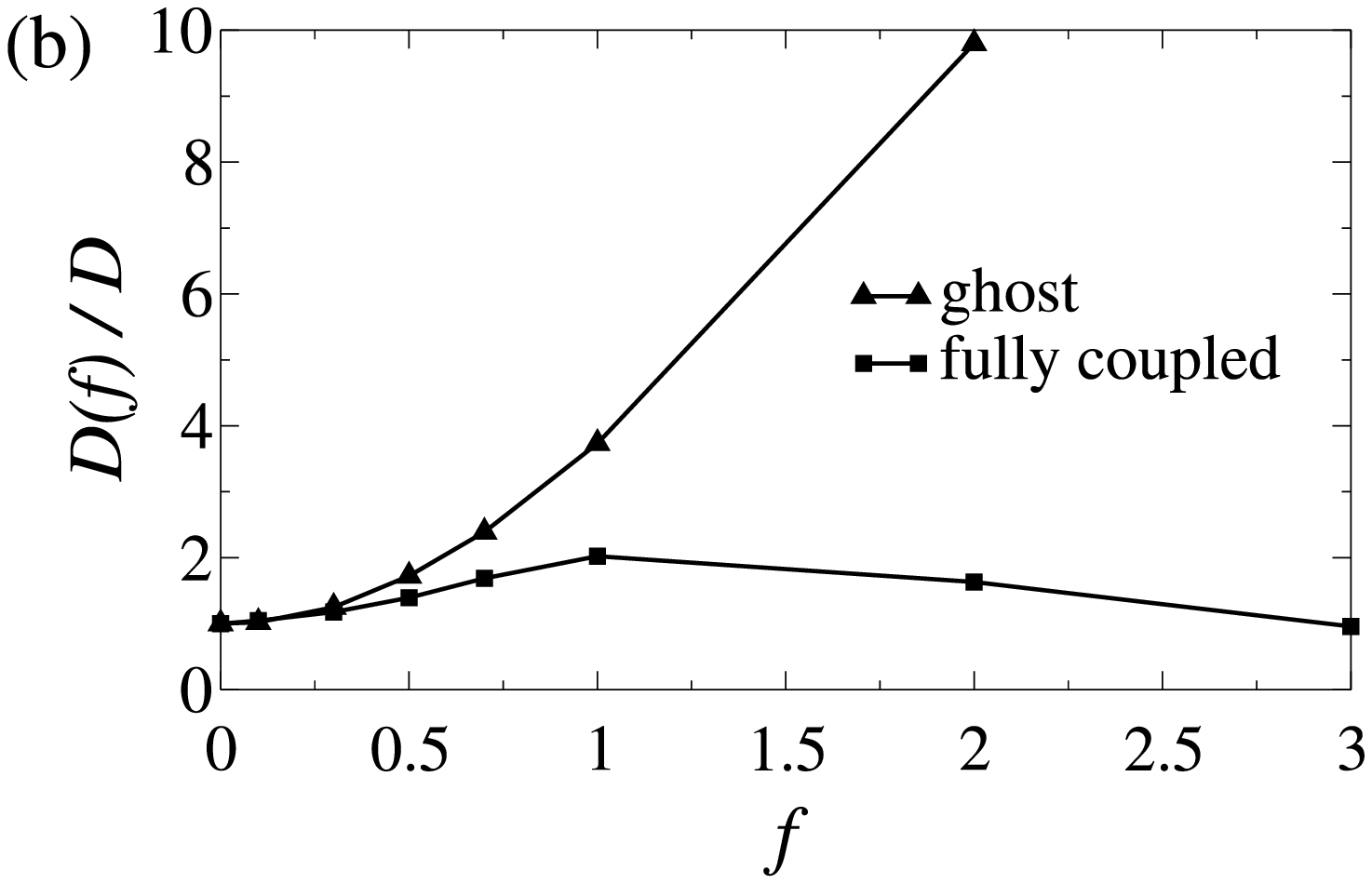}{8.5cm}{(Color online)
(a) Distribution of the probe displacement parallel
to the force,
$G(x,t)$, in the FA model at $\beta=5$ with `ghost' probes, showing
a bimodal structure.  The force is $f=0.5$ and the times are 
given in terms of the persistence
time $\tau=1.15\times10^6$ MC sweeps.  At the earliest time,
the peak at $x=0$ extends beyond the top of the figure.
(b) The force dependent diffusivity $D(f)$ at $\beta=3$ in the FA
model, 
comparing ghost probes (triangles) with fully coupled probes (squares).  
The force increases the diffusivity of the ghost probes. 
For the fully coupled probes, the mechanisms responsible
for the negative differential mobility also 
reduce the diffusivity at large forces.
}{fig:hist}{ht}

The analysis also allows us to estimate fluctuations around the average path.  
In particular, Eqs.~(\ref{equ:mu2}) and (\ref{equ:force-bias}) 
give the force-dependent diffusivity in $d=1$, 
\begin{equation}
D(f) = \frac{\sigma^2}{2\tau_\mathrm{x}} \left[ 1 + 2 \tanh^2(f/2) 
\left( \frac{\tau_\alpha}{\tau_\mathrm{x}}-1 \right) \right] .
\label{Dgiant}
\end{equation}
Equation~(\ref{Dgiant}) shows that increasing the force on the probe particle 
increases the diffusivity.  That is, it leads to larger 
fluctuations around the average path.  
For $d>1$, the functional form of $D(f)$ depends on the 
relationship between the force $\bm{f}$ and the bias $\bm{\delta}$,
but the qualitative picture remains the same as in $d=1$.  Moreover, 
the ratio $\tau_\alpha/\tau_\mathrm{x}$ is related to the Fickian length of~\cite{tauk}: 
\begin{equation}
\ell_\mathrm{F} \equiv 
\sqrt{D\tau_\alpha} = 
\sigma\sqrt{\frac{\tau_\alpha}{2d\tauxb}}.
\end{equation} 
In the deeply 
supercooled regime this length scale can
become large, $\ell_\mathrm{F} \gg \sigma$
\cite{tauk}, and the diffusivity $D(f)$ may increase
by orders of magnitude as the force is applied.

Fig.~\ref{fig:hist}(b) shows the increase of $D(f)$ with force 
in the FA model with `ghost' probes.   Even at the mildly 
supercooled conditions shown in the figure,
the diffusivity increases by at least an order of magnitude.  
A significantly smaller increase is found for the case of fully 
coupled probes.  The difference
between the two models arises because $\tau_\mathrm{x}$ increases with increasing force, Eq.~(\ref{tauxd}), so that the contribution of the second term in Eq.\ (\ref{Dgiant}) is suppressed.

As the relaxation slows in these systems, and fluctuations
grow, these non-linear fluctuation effects become important even at 
small forces.  From (\ref{Dgiant}),
we can identify the force at which the corrections to the
diffusivity become important as
$F_\mathrm{nl} \approx (k_\mathrm{B}T/\ell_\mathrm{F}) = 
(k_\mathrm{B}T/\sqrt{D\tau_\alpha})$.
Again, the Fickian length grows with decreasing temperature or 
increasing density, due to the increasing decoupling between 
diffusion and structural relaxation, and so $F_\mathrm{nl}$
decreases progressively as the system becomes more supercooled.  
A similar criterion was proposed in Ref.~\cite{Evans06} for the
onset of non-linear effects on probe velocities, 
$F_\mathrm{em} \approx (k_\mathrm{B}T/\sqrt{D t_\mathrm{c}})$, 
where $t_\mathrm{c}$ is a relaxation time beyond which certain probability
distributions converge to Gaussian forms.  The results of
that article also indicate a decrease of the threshold force
with decreasing temperature.

Finally, the intermittent motion of our forced particle leads to a two-peaked structure in the distribution of the probe displacement, as shown for the FA model in Fig.\ \ref{fig:hist}.  Similar distributions were observed in the atomistic simulations of Ref.~\cite{Evans06} and the experiments of Ref.\ \cite{grier}. 
They can be interpreted in our dynamical facilitation picture \cite{GC02, YJ05} in that the two peaks come from segregation of active populations and inactive populations.  Indeed, to a good approximation~\cite{tauk,convolution}, Eq.\ (\ref{MW}) gives
\begin{equation}
G(x,t) \approx P(t) \delta(x) + \frac{1-P(t)}{\sqrt{4\pi tD(f)}}
                         \exp\left\{-\frac{[x-tv(f)]^2}{4tD(f)}\right\} ,
\label{Gs}
\end{equation}
where $x$ is the displacement in the direction of the force. [Recall that the persistence function $P(t)$ is the fraction of probe particles that have not moved at all between time zero and time $t$.]  This equation for the distribution of probe displacements shows how the decoupling of exchange and persistence times
in dynamically heterogeneous materials~\cite{YJ05} leads 
to bimodal distributions such as those of Fig.~\ref{fig:hist},
and hence to the large fluctuations of the sort that in other contexts have been 
termed ``giant diffusivity''~\cite{grier}.

\section{Outlook}

We have shown in this article that weak forcing of probe particles in 
KCMs produces surprising non-linear responses such as negative
differential mobility and giant diffusivity. 
The origins of these non-linear effects 
are the heterogeneity \cite{YJ04,Ritort-Sollich} in the dynamics of the 
probe's host fluid; the decoupling between local exchange and persistence 
times \cite{YJ04,YJ05}; and the consequent intermittency \cite{YJ04,tauk} 
in the motion of the probe.  These same mechanisms naturally give rise to
transport decoupling in the absence of external forcing 
(for example, probe diffusion constants and structural relaxation times
have different scalings at low temperature
\cite{YJ04}).  
In comparison with other theoretical treatments of transport decoupling~\cite{SEother}, ours seems distinguished by the unifying connections it elucidates between the broad distributions of exchange and persistence times and various observed effects, now including negative response.

All our results are for low-dimensional KCMs.  We end with
a discussion of how they generalise to three dimensions, and
to atomistic or colloidal glass-formers.  
In explaining the fluctuation phenomena of Sec.~\ref{sec:fluct},
we assumed only that exchange and persistence times decouple
from one another.  This effect occurs in three-dimensional
KCMs in which fluctuations are large (for example, variants
of the KA or TLG models).  It was also recently demonstrated in 
three-dimensional atomistic glass-formers~\cite{Hedges}, and
so we expect giant diffusivity to be observed in those systems
also.  We have discussed how the results of Sec.~\ref{sec:fluct}
connect to previous atomistic simulations~\cite{Evans06}, and experiments 
would also seem feasible~\cite{grier}.  

In addition to decoupling of exchange and persistence times, 
negative differential mobility depends
on a suppression of the local exchange time by the 
applied force, as discussed in Sec.~\ref{sec:neg} and 
Sec.~\ref{sec:ctrw-neg}. 
This
feature seems to be generic in KCMs, and so we again expect
that it would generalise to three-dimensional KA and TLG models.
However, such a suppression of the exchange time 
has not yet been observed in three dimensional atomistic
or colloidal systems: we are not aware of experiments
that probe the relevant regime (recall Fig.~\ref{fig:vf_exp}).
As discussed in Ref.~\cite{Weeks04}, only the large force regime
is accessible using that system, since responses below
the threshold are smaller than the experimental resolution limits.
It seems that experiments with reduced forces $f\simeq1$ will
require the development of new methods.
However, searches for giant diffusivity and negative differential
mobility, in simulations of glass-formers
and in experiments, would provide a further test of the extent to 
which simple kinetically constrained models can be used to 
explain and predict the peculiar transport properties of
supercooled liquids.

\acknowledgments

In this research, JPG was supported by 
EPSRC under Grant No. GR/S54074/01.  DK was supported 
initially by the Director,
Office of Science, Office of Basic Energy Sciences, Chemical
Sciences, Geosciences, and Biosciences Division, U.S.
Department of Energy, under Contract No. DE-AC02-05CH11231.  RLJ and DC
were supported initially by NSF grant CHE-0543158 and later by Office of Naval
Research Grant No.~N00014-07-1-0689.


\begin{thebibliography}{99}
\bibitem{vel-footnote}
When calculating the value $v_0$ for the experimental data,
we use results for
the colloid particles of~\cite{Weeks02} at volume fractions $0.53-0.56$,
since the very small diffusion constants for probe particles 
are not accurately determined in Ref.~\cite{Weeks04}.  
Possible differences in the values
of $v_0$ for probes and colloids should make only small differences
on the logarithmic scale used in Fig.~\ref{fig:vf_exp}.
\bibitem{inviolable-foot}
It would be interesting to investigate the large-force
regime by considering KCMs in which constraints can be
violated with force-dependent rates.  However, while
KCMs capture the steric `crowding' effects associated
with glassy behavior, details of the interparticle
interactions have been coarse-grained away, as have interactions
between the particles and the solvent.  For these reasons,
the development of a lattice model that accurately
describes the colloidal system for forces above the
threshold is beyond the scope of this paper, and we
restrict ourselves to the regimes of
small and intermediate force, shown in Fig.~\ref{fig:vf_exp}.
\bibitem{latt-foot}
For diffusion in a continuous medium, the only relevant parameter
is the diffusion constant $D$.  In discretising this medium onto
a particular lattice with spacing $\sigma$, 
the rate for (unbiased) hops $\gamma$ is determined by the diffusion
constant.  If Glauber dynamics are used the maximal probe
velocity is $\sigma\gamma\propto(D/\sigma)$.  Thus, for a given
diffusion constant, the
maximal probe velocity is finite in the lattice system, but
the saturation disappears in the continuum limit $\sigma\to0$.
In this sense, the saturation is a lattice artifact.
\bibitem{East-foot}
We note that the dynamical rules of the East model are 
not invariant under spatial reflection. 
However, in the absence of an
external field, time reversal symmetry ensures that
the probe motion is symmetric in space.
It would be interesting to study the
interplay between the intrinsic asymmetry of this model and
the asymmetric perturbation introduced by the force.
In particular, the response of probe particles may depend 
on the direction of the applied force.
\bibitem{Ghost}
We have found a similar distinction between ghost and fully
coupled probes in other facilitated models, 
such as Fredrickson and Andersen's two-spin facilitated model in dimensions 
two and higher.
\bibitem{alpha-foot}
While this correspondence is standard in unforced 
systems~\cite{Ritort-Sollich,GC02,YJ05}, 
the probe persistence time might depend
on the applied force.  However, while the force-dependence of the
exchange time has important consequences (see Sec.~\ref{sec:ctrw-neg}),
a weak force-dependence of the persistence time has no qualitative
effect on our results, so we neglect it.
\bibitem{convolution} The inverse Laplace transform of Eq.~(\ref{MW}) gives 
for the second term of Eq.~(\ref{Gs}) a convolution 
between the persistence time distribution and the diffusion propagator.  
The second term in Eq.~(\ref{Gs}) is an approximation to that 
convolution whose long time limit is consistent with 
Eqs.~(\ref{rs}-\ref{equ:mu2}), and
which captures the crossover from the immobile to the mobile regime.  
The crossover regime has been considered explicitly in 
Ref.~\cite{chaudhuri}.
\end{thebibliography}
\end{document}